\documentclass{skaox2006}
\usepackage{graphicx}
\newcommand{\Gaia}{\emph{Gaia}}
\begin{document}
   \title{The European Space Agency {\Gaia} mission: exploring the Galaxy}

   \author{C. Jordi}

   \institute{Departament d'Astronomia i Meteorologia, Institut de Ci\`encies
              del Cosmos, Universitat de Barcelona--ICC/IEEC, c/ Mart\'{\i} 
              i Franqu\`es, 1, 08028 Barcelona, Spain}

   \abstract{

The {\Gaia} astrometric mission was approved by the European Space Agency 
in 2000 and the construction of the spacecraft and payload is on-going for a
launch in late 2012. 
{\Gaia} will continuously scan the entire 
sky for 5 years, yielding positional and velocity measurements with the 
accuracies needed to produce a stereoscopic and kinematic census of about 
one billion stars throughout our Galaxy and beyond.
The main scientific goal is to quantify 
early formation and the subsequent dynamic and chemical evolution of the Milky way.
The stellar survey 
will have a completeness to $V = 20$~mag, with a precision of about 25~$\mu$as 
at 15~mag. The astrometric information will be combined with astrophysical 
data acquired through on-board spectrophotometry and spectroscopy, 
allowing the chemical composition and age of the stars to be derived. 
Data acquired and processed as a result of the {\Gaia} mission 
are estimated to amount to about 1 petabyte. One of the challenging problems is 
the close relationship between astrometric and astrophysical data, which 
involves a global iterative solution that updates instruments parameters, 
the attitude of the satellite, and the properties of the observed objects. 
The European community is organized
to deal with {\Gaia} products: (a) the Data Processing and Analysis Consortium is
a joint European effort in charge of preparation and execution of data
processing, (b) the GREAT network is a platform for collaboration on the
preparation of scientific exploitation.

}
   \maketitle
%
%
\section{Introduction}

The {\Gaia} mission aims to study the origin, formation, and evolution of the 
Milky Way and its components. Our Galaxy mainly contains 
stars of several types and ages, interstellar gas and dust in a disk, a spheroidal 
halo with old stars, a bar and a 
bulge in the center, and the ubiquitous dark matter. The disk shows a 
spiral structure, with the arms being the preferred location of the 
star-forming regions. Although these general features are rather well 
known, much remains to be elucidated, including the number and detailed
structure of the spiral arms, the disk warp, the shape and 
rotation of the bulge and halo, the global dynamics and the kinematics, and the distribution of dark matter. Also, the process in 
which the Galaxy was assembled, presumably from small building blocks, and 
the history of star formation are not well understood.

To address these, one needs a deep full-sky survey covering a significant 
volume of the Galaxy. This is what {\Gaia} will do (ESA \cite{ESA2000}, 
Perryman et al \cite{ESA2001}) by 
creating the largest and most precise three-dimensional survey of our Galaxy 
and beyond with unprecedented positional, proper motion, radial 
velocity, and spectroscopic data for about one billion stars in our Galaxy 
and throughout the Local Group. The {\Gaia} launch is scheduled for late
2012 and the mission ends 5 years later. The detailed design and 
construction phase started in February 2006, with the development and 
construction are progressing as planned.

The operating principle is that of its 
successful predecessor, the Hipparcos satellite of the European Space 
Agency (ESA) (ESA \cite{ESA1997}, Lindegren et al \cite{Lennart}). 
{\Gaia} performs wide-angle 
astrometry by means of two telescopes and two associated viewing 
directions, allowing the determination of absolute trigonometric 
parallaxes. The size of the primary mirrors and of the field of views and 
the use of CCD detectors in the focal plane mean an enormous step further 
than achieved with the Hipparcos mission. While Hipparcos yielded 
precisions of 1 mas at 9 mag and measured about 120,000 stars, {\Gaia} will 
yield 25 $\mu$as at 15 mag and will survey a billion objects. Although 
Hipparcos seems, in retrospect, like a modest project, it was the first 
space astrometric mission and it revolutionized our view of the Galaxy, despite the reduced number of obseved stars. 
{\Gaia} builds on the 
experience acquired with Hipparcos; it is an entirely European 
mission supported by a wide and motivated scientific and technological 
community, which holds the leadership in the 
space astrometry domain.

This review discusses the scientific goals of the mission and the mission 
itself, with a description of its operating principle and instruments, and
the status of development as 
well as the organizational scheme with respect to its bodies and 
responsibilities.  A detailed presentation of the scientific 
case and mission design at the 2000 stage can be found in ESA (\cite{ESA2000}) and a review of the mission at the 2004 stage in Turon \& Perryman (\cite{turon}).


\section{The scientific case}

The scientific goals of {\Gaia} have been extensively discussed (ESA \cite{ESA2000}, 
Perryman et al \cite{ESA2001}), the 
main one being the production of a homogeneous and accurate catalogue 
to carry out the deepest and most well grounded study thus 
far of the structure, origin, formation, and evolution of the Milky Way. 
Since {\Gaia} is a survey mission, the 
observations are not limited to the stars in the Galaxy, since any 
object in the fields of view will be recorded. Thus, {\Gaia} will 
impact on all fields of astronomy and astrophysics as well as on fundamental 
physics. In the following, we outline some of 
science goals.

\begin{itemize}
\item{{\em Galactic structure, kinematics and dynamics and star-formation history}} 

The huge number of stars to be observed by {\Gaia}, the impressive astrometric accuracy, and the faint limiting magnitude will allow us to 
quantify the structure and motions of stars within the bulge, the spiral 
arms, the disk, and the outer halo. The positions and velocities
are linked through gravitational forces, and through the star-formation 
rate as a function of position and time. The initial distributions are 
modified, perhaps substantially, by small- and large-scale dynamic 
processes, including instabilities that convey angular momentum and mergers with other galaxies. 

The star-formation history defines the luminosity evolution of 
the Galaxy directly. In combination with the relevant chemical abundance 
distributions, the accretion history of gas may be derived, while together 
with kinematics, the merger history of smaller stellar systems can be 
defined. The sum of these three processes comprises what is loosely known 
as “galaxy formation,” and the {\Gaia} mission 
will provide the first quantitative determination of the formation history 
of our Galaxy.

\item{{\em Stellar structure and evolution}}

  The study of stellar structure and evolution provides fundamental
  information on the properties of matter under extreme physical
  conditions.  {\Gaia} will return information about luminosities,
  surface temperatures, abundances, masses for all types of stars,
  including rare objects. This will allow us to study the sizes of the
  convective cores of massive stars, the internal diffusion of the
  chemical elements, and the outer convective zones and to solve the
  current discrepancies among observations and theory.  In addition,
  {\Gaia} will provide multi-epoch, multi-color photometry. These data
  will permit detection of diverse variable phenomena, allowing a
  global description of stellar stability and variability across the
  Hertzsprung-Russell diagram.

\item{{\em Stellar ages and age of the universe}}

  The primary age-determination method relies on comparisons of
  stellar models or isochrones with the best available data for
  individual stars or stellar groups. Determination of the age of the
  oldest stars provides a lower limit to the age of the universe,
  which in turn constrains cosmological models.  {\Gaia} will improve
  the age estimate of the oldest stars by improving their
  distance--luminosity determination.  The number of subdwarfs with
  accurate distances will considerably increase in each metallicity
  interval, allowing us to derive the distance of a greater number of
  globular clusters of various chemical compositions.

\item{{\em Distance scale}}

Measures of trigonometric parallaxes will be unique. {\Gaia} will have a 
major impact upon our knowledge of the distance scale in the universe by 
providing accurate distances and physical parameters for all types of 
observable primary distance indicators in the Milky Way and in the closest 
galaxies of the Local Group. It will generate a complete sampling of these 
indicators, allowing the corrections necessitated by metal, oxygen, or 
helium contents, color, population, age, etc. In particular, {\Gaia} will 
observe countless 
Cepheids and RR Lyrae, thus providing solid calibrations for 
cluster-sequence fitting and period-luminosity relations.

\item{{\em Binaries, multiple systems and planetary systems}}

One of {\Gaia}'s unique features is the unbiased survey over the entire 
sky. In looking at a nearby sample, many resolved binaries 
have periods short enough for orbit determination by {\Gaia}. The study of 
binaries will facilitate the discovery of thousands of low-mass 
low-luminosity companions of stellar nature, brown dwarfs, or planets
through the motion inferred to their parent stars.
The detection of 10,000--20,000 exoplanets and the 
determination of orbits for some 5000 of them is expected, with estimates of the actual planet masses.

\item{{\em Solar system}}

It is estimated that some $10^5-10^6$ asteroids will be 
discovered (compared with the 65,000 known), plus some hundreds of 
Kuiper's belt objects. Orbits of Near-Earth objects will be 
known with a precision 30 times higher than 
what is possible today. Positional and photometric observations spanning 
the 5-year mission will provide masses, sizes, and composition, as a function of the distance to the Sun, thus improving our 
understanding of the history of the solar system's formation.

\item{{\em Gravitational light deflection}} 

The detected photons are bent during the last hours of their long journey, 
under the influence of the gravitational fields of the Sun, 
planets, moons, asteroids, etc. The amount of this light 
deflection depends on the mass of the perturbing object, its distance to 
the observer, and the angular separation at which the photon passes 
the object. The 
actual measurements of the blending of the light will constrain the 
various parameters of post-Newtonian theory.

\item{{\em Extragalactic astrophysics}}

  Astrometric and photometric observations for about 500,000 active
  galactic nuclei and quasars will be derived over the whole sky,
  allowing us to establish the inertial reference frame to an accuracy
  of $\sim$0.4 $\mu$as yr$^{-1}$. Galactocentric acceleration of the
  Sun will be derived at a precision of 0.2 nm s$^{-2}$. It is
  expected that $\sim$100,000 supernovae will be discovered, and
  $\sim40\cdot10^6$ galaxies will be surveyed.

\end{itemize}

\section{{\Gaia} operations, instruments and status of development}

\begin{figure}[h]
   \centering
      \includegraphics[scale=0.8]{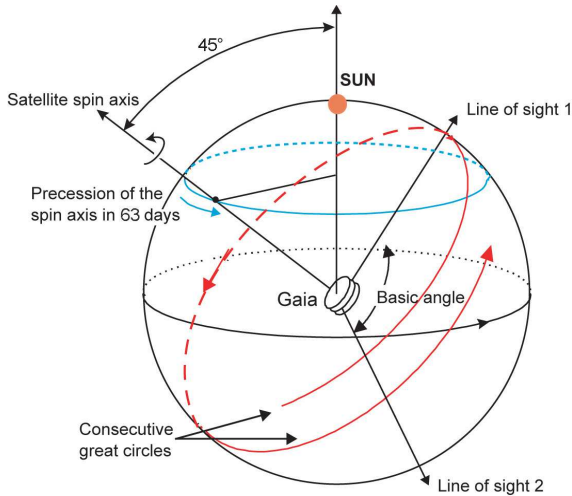}
   \caption{{\Gaia}'s two astrometric fields of view scan the sky according 
to a carefully prescribed revolving scanning law. The constant spin rate 
of 60 arcsec s$^{-1}$ corresponds to 6-hour great-circle scans. The basic 
angle between the two lines of sight is 106.5~deg. (Courtesy of Karen O'Flaherty).
            \label{fig:principle}
           }
    \end{figure}

Operating from a Lissajous orbit around the second Lagrange point of the 
Sun-Earth/Moon system, the satellite will continuously scan the sky. 
During its 5 years of life, {\Gaia} will rotate, at a fixed speed of 
60 arcsec s$^{-1}$, around a slowly precessing spin axis at 45~deg of the Sun (Fig.~\ref{fig:principle}). 
As a result, objects continuously traverse the focal planes. {\Gaia}  
has two telescopes (focal length is 
35~m) with two associated viewing directions. The primary 
mirrors are $1.45\times 0.5$~ m$^2$ in size each. The two viewing angles are separated by a highly stable basic angle 
of 106.5~deg.

\begin{figure*}
   \centering
   \includegraphics[scale=0.20]{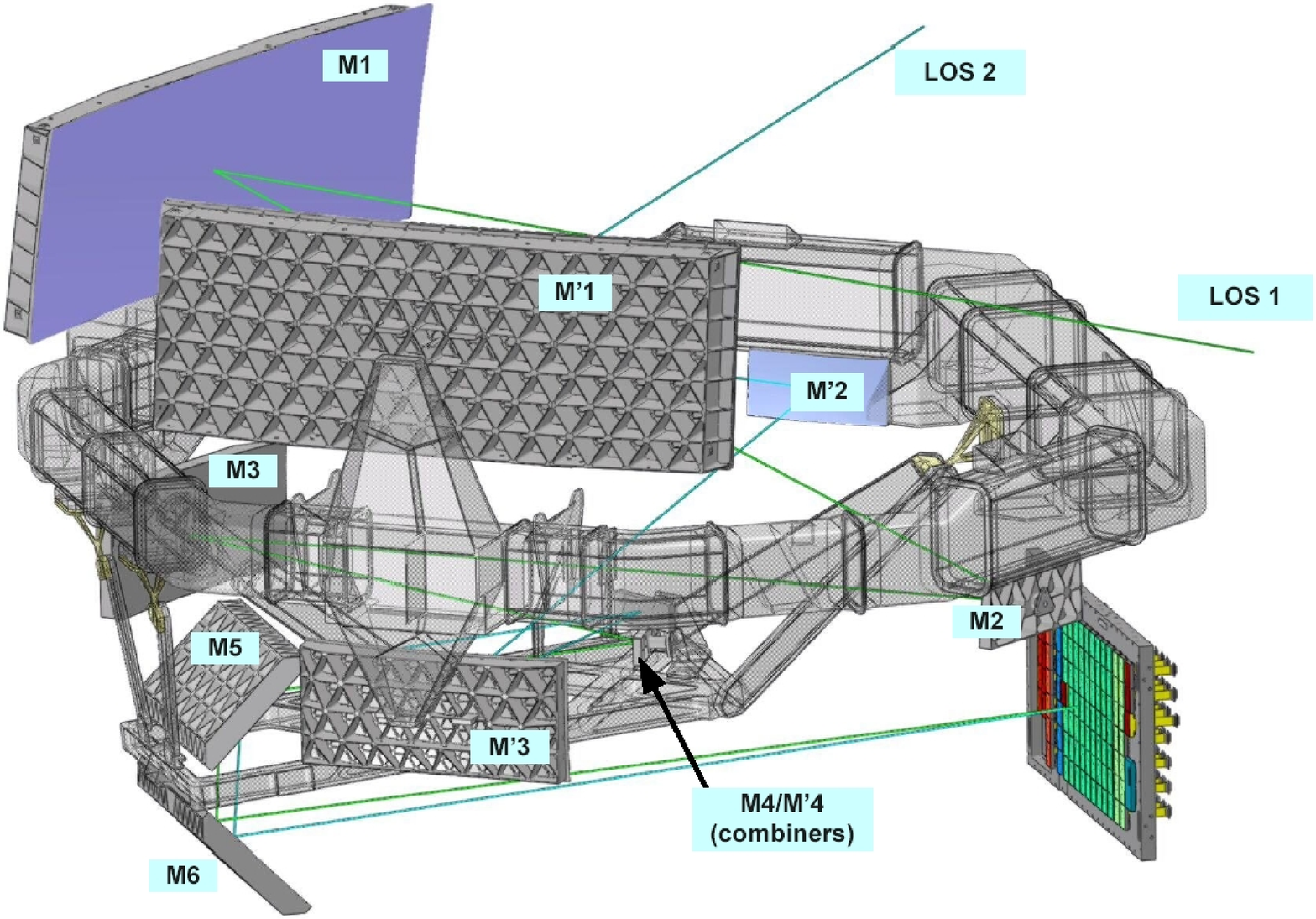}
   \includegraphics[scale=0.205]{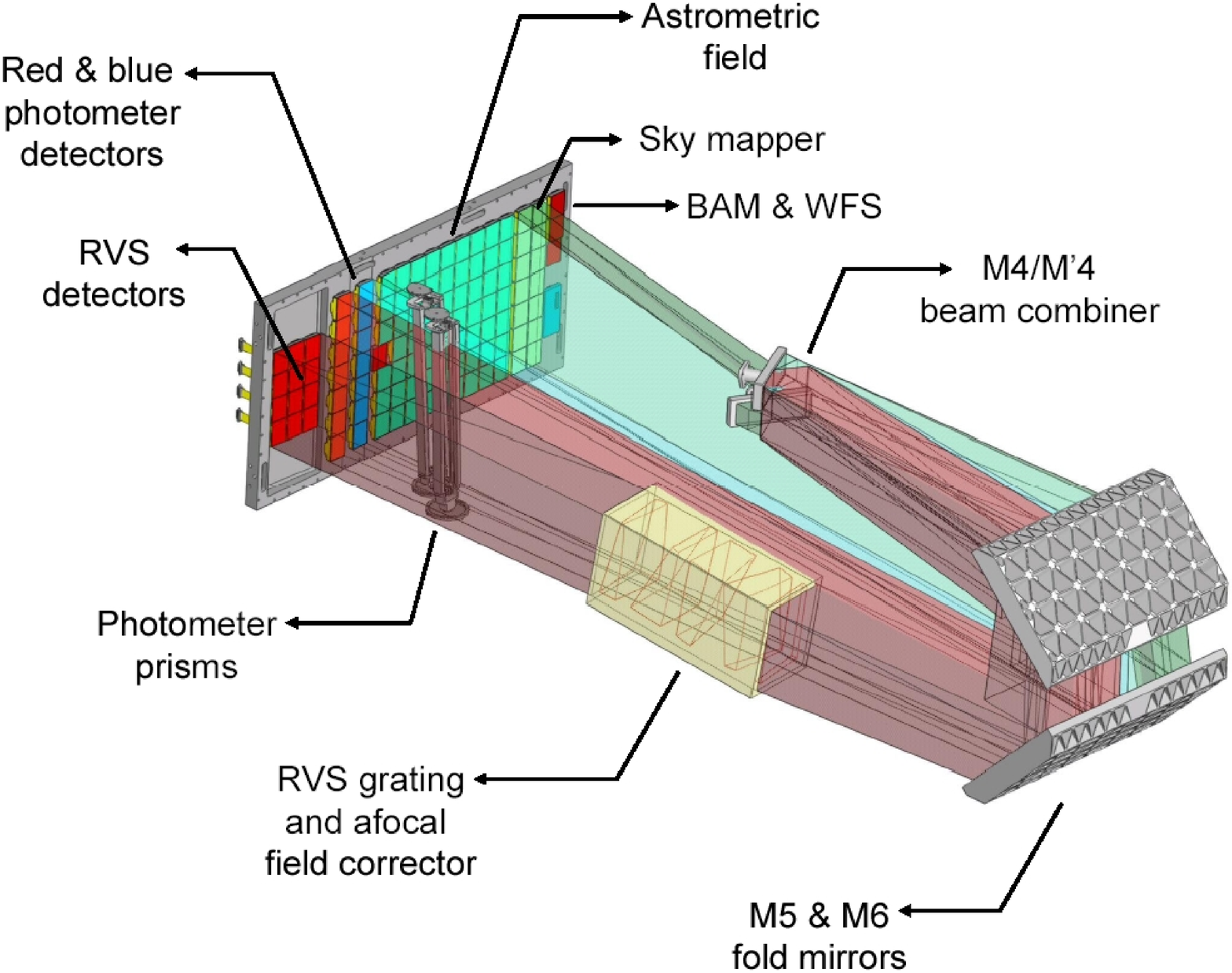}
   \caption{(Left) {\Gaia} payload and telescopes. LOS1\&2 are the two lines of 
sight, M1\&2 the two primary mirrors, and M2-6 the mirrors that drive the 
image to the combined focal plane. (Right) The configuration of the slitless spectrograph with 
the blue and red prisms and of the radial-velocity grating combined with the 
focal plane and the M4--6 mirrors of the telescopes. (Courtesy of EADS-Astrium)
            \label{fig:telescope}
           }
    \end{figure*}

    The telescope assemblies consist of four mirrors for each line of
    sight and two common additional mirrors
    (Fig.~\ref{fig:telescope}). The two fields of view are combined
    into a single focal plane of $104 \times 42$ cm$^2$ (along scan
    $\times$ across scan) covered with 106 CCD detectors. All the
    parts are mounted on a torus and all of them are made of SiC,
    offering thermal and mechanical stability. The mirrors are silver
    coated.

   \begin{figure}
   \centering
   \includegraphics[scale=1.1]{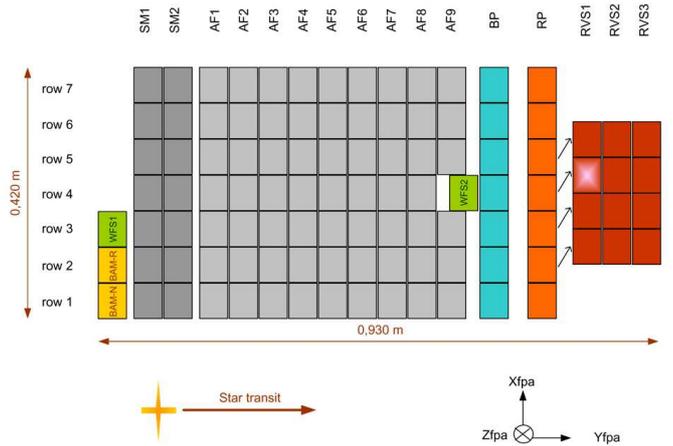}
   \caption{Focal plane of {\Gaia}. SM1\&2 will detect objects 
at field of view 1 and 2, respectively. AF1--9 are the CCDs collecting
unfiltered light for astrometry. BP \& RP are for blue and red
spectrophotometers, respectively. RVS1--3 are for spectroscopy. BAM and WFS are
the basic-angle monitors and wave-front error sensors, respectively. Stars
transit from left to right as the satellite scans the sky. (Courtesy of J. de
Bruijne).
            \label{fig:focalplane}
           }
    \end{figure}

{\Gaia} observations will be made using high-quality, large-format CCDs 
in the common focal plane (Fig.~\ref{fig:focalplane}). The CCDs are operated in time delay 
and integration mode. Charge images will be transported  in 
synchrony with images moving across the field due to the rotation 
of the satellite. The focal plane has four main sections:

a. Stars entering the focal plane first pass across dedicated 
CCDs that act as a “sky mapper” (SM). Here, the objects are detected 
and the strategy of observations in the rest of the focal plane is 
established case by case (only pixels around the objects are 
read and sent to the ground).

b. The astrometric field (AF) is sampled by an array of 62 CCDs. 
Astrometric observations are made with unfiltered light in order to 
minimize photon noise. The mirror coatings and the CCD QE effectively 
define a broad (white-light) passband covering the wavelength range of 
about 350--1000 nm, with a mean wavelength of 673~nm and a
full width at half-maximum of 440~nm.

c. The spectrophotometric CCDs record low-resolution spectra produced by 
two slitless spectrographs, BP (blue photometer) and RP (red photometer), 
covering the wavelength intervals 330--680~nm and 650--1050~nm, 
respectively. The goal is to provide astrophysical classification and 
parameterization of observed objects.

d. The radial velocity spectrometer (RVS) CCDs, combined with 
high-resolution integral field spectrograph ($R = 11500$) in the range 
847--874~nm on the IR Calcium triplet, allows derivation of the radial 
velocities and chemical composition of the brightest stars.

By measuring the instantaneous image centroids from the data sent to 
ground, {\Gaia} measures the relative separations of the thousands of stars 
simultaneously present in the combined fields. Scans in different 
directions during the 5-year mission allow the measurement of a given star 
in relation to many others in differently oriented great circles. At the 
end of the mission, this translates into a precision in angle 
measurements of about 25~$\mu$as at $V = 15$, that is, 25\% precision in 
distance at 10~kpc, and equivalent to measuring a Euro coin at the 
Moon distance. This kind of measurement also allows the separation of 
parallax-induced motion in the sky from the star's own motion, thus 
yielding the distance and the proper motion of the star with respect to 
the observer. In summary, the targeted numbers at the end of the mission 
are:
\begin{itemize}
\item One billion objects ($0.34\cdot10^6$ to $V = 10$ mag; $26\cdot10^6$ 
to $V = 15$ mag; $250\cdot10^6$ to $V = 18$ mag)

\item Positions and proper motions with precisions better than 25 $\mu$as and 
25 $\mu$as yr$^{-1}$ at $V = 15$

\item Parallax data with 25 $\mu$as at $V = 15$

\item Stellar atmospheric parameters (temperature, gravity, chemical 
composition) for all stars up to $V = 18$

\item Radial velocities of $\sim15$ km s$^{-1}$ precision at $V = 17$
\end{itemize}

The total mass of the satellite including payload, service module, etc., 
amounts to 2100 kg, requiring a power of 1630 W that is to be provided by 
an assembly of solar panels. The spacecraft will be launched by a 
Soyuz-Fregat launcher from Kourou facilities. The ground-segment is 
provided by the 35-m antenna of ESA's Cebreros station, with a rate of 
30 GB per day downlink. When {\Gaia} is scanning the galactic plane, a second 
antenna, at the New Norcia station, will provide the necessary additional 
telemetry time.

\begin{figure}
   \centering
   \includegraphics[scale=0.60]{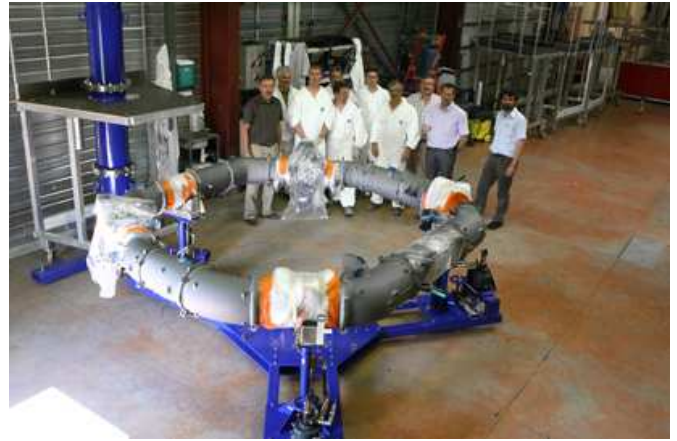}
   \caption{Members of the BOOSTEC and Astrium teams just after the torus
removal from the brazing furnace. The torus is made of SiC and consists of 16
different pieces independently manufactured. (Courtesy of ESA).
\label{fig:torus}
           }
    \end{figure}

\begin{figure}
   \centering
   \includegraphics[scale=0.3]{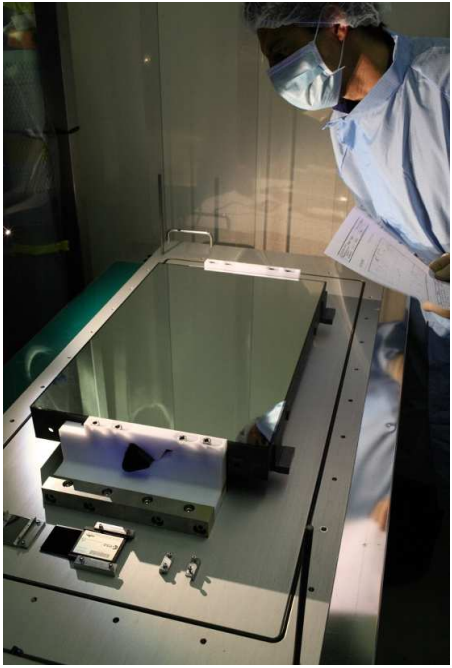}
   \caption{The $540 \times 336$~mm M5 mirror undergoing inspection at the
premises of Advanced Mechanical and Optical Systems (AMOS) in Liège, Belgium,
in November 2009. The SiC structure was made by Boostec Tarbes, France; SiC
vapor deposition on the surface was carried out by Schunk Kohlenstofftechnik at
Heuchelheim, Germany. (Courtesy of ESA).  \label{fig:mirror}
           }
    \end{figure}

The industrial consortium, selected in February 2006, for the {\Gaia}  
spacecraft is spread over Europe, with EADS-Astrium at Toulouse acting as 
prime contractor. The design, building, and test phase is currently 
ongoing, with expected finalization in mid 2012. The payload has passed 
the critical design review and the spacecraft critical design review is 
foreseen during summer 2010. The sixteen segments composing the torus 
structure were built and the torus has been successfully brazed (Fig.~\ref{fig:torus}), 
one of the main achievements of the building phase. The SiC substrate 
structure of the mirrors has been manufactured and polished. Testing and 
coating are ongoing for some of them, and some have already been coated, 
finalized, and delivered (Fig.~\ref{fig:mirror}). The delivery of all 
mirrors shall be
completed by summer 2010. CCD production is also in progress, 
with 85\% of the devices already finalized and tested. Other elements of
the scientific instruments and spacecraft are either finalized or under
planned development. 


One of the main critical items is the charge inefficiency that occurs when 
moving the charge under the operating conditions of radiation damage to 
the CCDs by proton particles from the solar wind. This charge inefficiency 
translates into delayed transport and a severe deformation of the 
diffraction images on the focal plane. Tests are currently being conducted 
to characterize on ground the several dependencies of deformations. These 
dependencies wll be used as input knowledge to the a posteriori on-ground 
treatment of real {\Gaia} data.


\section{Organizational structure and responsibilities}

The {\Gaia} satellite and mission operation is fully funded by ESA. The 
management structure includes a project 
manager and a project scientist. The project manager is in charge of 
supervising industrial development and ground segment management. The 
project scientist supervises the accomplishment of the scientific goals in 
the design phase and is assisted by a group of scientists external to ESA 
and representing the scientific community. 

Processing of the acquired data is the responsibility of the 
Data Processing and Analysis Consortium (DPAC, Mignard et al \cite{dpac}), comprising more than 
430 members in 11 countries who are joining their efforts to overcome the 
challenging problem of managing the 1 petabyte of {\Gaia} data. DPAC 
activities are supported by governments that have signed a multilateral 
agreement with ESA to ensure the stability of the scientific teams for the 
necessary interval of time.
The work inside DPAC is presently organized in eight coordination units, with
fields of competence and leaders. 
Processing will be done on the premises of six Data Processing
Centers also made up of scientists and engineers from the countries who
have signed the multilateral agreement. 

Linked to DPAC, the Marie Curie Research Training Network ''European Leadership on Space Astrometry (ELSA)'' was created. Its aim
is the development and training of the next generation of experts in the field
of astrometry. 

Neither ESA nor DPAC has direct responsibility for {\Gaia} data exploitation,
which, instead, is to be conducted by the scientific community at large. It is
up to this community to organize the best use of {\Gaia}'s products. Since the
DPAC community has deep knowledge of the {\Gaia} mission and extensive expertise
in astrometry, synergy with the community at large is obvious and necessary for
making use of the enormous legacy that {\Gaia} will provide. To that end, several
networks are currently in place, such as GREAT ({\Gaia} Research for European
Astronomy Training), founded by the European Science Foundation and others at national level. These networks aim to strengthen the
cooperation between teams, thereby joining expertise in observations, {\Gaia}  
knowledge, theoretical analysis, model development and interpretation, and
statistical treatment. Complementary observations on-ground covering the areas
of science alerts, IR photometry, and high resolution spectroscopic surveys are
being studied. 


\section{Summary}

The {\Gaia} mission is one of the cornerstone missions within the cosmic 
vision of ESA's program. 

The development and manufacturing phase is well on track towards a 
scheduled launch in August 2012. Transfer to orbit around the second Lagrange 
point and commissioning will take up to 3 months, to be followed by the 
routine science operations phase. The science operations will last 5 
years, with a possible extension of one year. Data processing will start 
as soon as data are received by the on-ground segment and will continue 
some 2--3 years after the mission's end. The final catalogue is 
expected around 2021, but with intermediate data releases produced during 
the operational phase.

A scientific and technical community of more than 400 members is currently 
involved in this project, which maintains European leadership in the space 
astrometry field. The scientific community is excited by {\Gaia}'s goals 
and products and by the legacy that it will represent for future 
generations of astronomers. Doubtless, {\Gaia} will revolutionize the concept 
of our Galaxy and beyond and the view of our solar system.

\begin{acknowledgements}
The project is funded by the Spanish MICINN under contract 
AYA2009-14648-C02-01.
\end{acknowledgements}

\end{document}